\documentclass[a4paper,11pt]{article}
\pdfoutput=1 
\usepackage{jheppub} 
\usepackage{slashed}
\usepackage[T1]{fontenc} 
\usepackage{graphicx}
\usepackage{epsfig}
\usepackage{amsmath,amssymb}
\usepackage{MnSymbol}
\usepackage{physics}
\usepackage{esint}

\def\be{\begin{equation}}
\def\ee{\end{equation}}
\def\ba{\begin{aligned}}
\def\ea{\end{aligned}}
\def\p{\partial}

\def\sig{\sigma}

\def\half{\frac{1}{2}}

\numberwithin{equation}{section}

\title{Hidden Symmetry in the Double Copy}

\author{Adam Ball,}
\author{Anna Bencke,}
\author{Yaxi Chen,}
\author{and Anastasia Volovich}

\affiliation{Department of Physics, Brown University, Providence, RI 02912, USA}

\emailAdd{adam\_ball@brown.edu,
anna\_bencke@brown.edu, yaxi\_chen@brown.edu, anastasia\_volovich@brown.edu}

\abstract{We show that the Killing tensor of the Kerr spacetime has an analogue in the $\sqrt{\rm Kerr}$ gauge theory solution related to it by the classical double copy. This hidden symmetry of $\sqrt{\rm Kerr}$ leads to an additional constant of motion for color-charged point particles moving in it, implying integrability of the equation of motion. These are the gauge theory counterparts to the Carter constant and the integrability of the geodesic equation in a Kerr background.}

\begin{document} 
\maketitle
\flushbottom


\section{Introduction}

The double copy relation between gravity and gauge theory was originally discovered in the context of perturbative scattering amplitudes \cite{Bern:2008qj, Bern:2010yg, Bern:2010ue}, but it was soon shown in \cite{Monteiro:2014cda, Luna:2015paa} that a version of it also applies to certain classical solutions of gravity and gauge theory. This classical double copy has since been extended and clarified substantially, e.g. in \cite{Ridgway:2015fdl, Luna:2016due, White:2016jzc, Luna:2016hge, DeSmet:2017rve, Carrillo-Gonzalez:2017iyj, Ilderton:2018lsf, Berman:2018hwd, Gurses:2018ckx, Luna:2018dpt, CarrilloGonzalez:2019gof, Bah:2019sda, Huang:2019cja, Alawadhi:2019urr, Kim:2019jwm, Keeler:2020rcv, Elor:2020nqe, Alawadhi:2020jrv, Easson:2020esh, Monteiro:2020plf, Alkac:2021bav, Chacon:2021hfe, Easson:2021asd, Luna:2022dxo, Chawla:2022ogv, Easson:2022zoh, Comberiati:2022cpm, Chawla:2023bsu, Easson:2023dbk}, but its fundamental underpinnings remain somewhat elusive and it still holds mysteries. One direction of generalization was described in \cite{Gonzo:2021drq}, where the authors established a version of the double copy for point particles moving in double-copy-related backgrounds. They expounded in particular on the Kerr solution to gravity and the so-called $\sqrt{\rm Kerr}$ solution to Yang-Mills theory. Here we build on their work, allowing for non-equatorial trajectories and translating the results about particle motion back to results about the classical double copy itself. We find that the hidden symmetry of the Kerr solution is mapped by the double copy relation to a hidden symmetry of the $\sqrt{\rm Kerr}$ solution.

In section \ref{sec:KSreview} we review the Kerr-Schild double copy. In section \ref{sec:GonzoReview} we review the results of \cite{Gonzo:2021drq}. In section \ref{sec:CarterReview} we review the Carter constant and corresponding integrability of the Kerr solution. In section \ref{sec:HidSym} we present our results, namely the gauge analogue of the Carter constant, the reduction of particle motion in the $\sqrt{\rm Kerr}$ solution to quadratures, and the gauge analogue of the Kerr solution's Killing tensor. Finally in section \ref{sec:Discussion} we further contextualize our results and discuss prospects for generalization.

\section{Review of Kerr-Schild Double Copy}
\label{sec:KSreview}

A surprisingly wide class of metrics can be written in Kerr-Schild form (see e.g. \cite{Stephani:2003tm} for a review),
\be \ba g_{\mu\nu} & = \bar g_{\mu\nu} + h_{\mu\nu} \\
& = \bar g_{\mu\nu} + \varphi k_\mu k_\nu, \ea \ee
where $\bar g_{\mu\nu}$ is the flat metric and the covector $k_\mu$ is null and geodesic with respect to it, satisfying
\be k_\mu \bar g^{\mu\nu} \bar\nabla_\nu k_\rho = 0 \ee
with $\bar\nabla_\mu$ the covariant derivative of $\bar g_{\mu\nu}$. This property fixes $\varphi$ up to a constant. The inverse metric takes the simple form
\be g^{\mu\nu} = \bar g^{\mu\nu} - \varphi k^\mu k^\nu, \ee
where the indices on the right hand side have been raised with $\bar g^{\mu\nu}$. Note that $k_\mu$ is null with respect to $g_{\mu\nu}$ as well as $\bar g_{\mu\nu}$, and its index can be raised equally well with either metric. One can also show that $k_\mu$ is geodesic with respect to $g_{\mu\nu}$.

The Kerr-Schild double copy \cite{Monteiro:2014cda, Luna:2015paa} states that if $h_{\mu\nu} = \varphi k_\mu k_\nu$ is a stationary Kerr-Schild perturbation, then $A_\mu^a = \varphi k_\mu \tilde c^a$ solves the Yang-Mills equations in the flat background $\bar g_{\mu\nu}$ for any constant color vector $\tilde c^a$. This relationship,
\be \label{eq:dblcopy} h_{\mu\nu} = \varphi k_\mu k_\nu \quad \longleftrightarrow \quad A_\mu^a = \varphi k_\mu \tilde c^a, \ee
is pithily summarized as
\be \label{eq:pith} k_\nu \,\, \longleftrightarrow \,\, \tilde c^a, \ee
which states a sort of duality between kinematics and color \cite{Bern:2008qj, Bern:2010yg, Bern:2010ue}. Note that since the color behavior in $A_\mu^a$ is just a constant factor $\tilde c^a$, the Yang-Mills equations linearize and the gauge field is effectively abelian.

Our focus in this paper is on the Kerr metric \cite{Kerr:1963ud}, which in Kerr-Schild form in Cartesian coordinates $(t, x, y, z)$ is \cite{dInverno:1992gxs}
\be g_{\mu\nu} = \bar g_{\mu\nu} + \varphi k_\mu k_\nu \ee
with
\be \varphi = \frac{2GM r^3}{r^4 + a^2 z^2} \ee
and
\be k_\mu = \left( 1, \frac{rx+ay}{r^2+a^2}, \frac{ry-ax}{r^2+a^2}, \frac{z}{r} \right). \ee
Here $r$ is defined implicitly by
\be 1 = \frac{x^2 + y^2}{r^2 + a^2} + \frac{z^2}{r^2}. \ee
The black hole's mass is $M$ and its angular momentum is $aM$. The particular value of the mass will be irrelevant to us, so henceforth we set $2GM = 1$. The corresponding gauge theory solution, which we refer to as the $\sqrt{\rm Kerr}$ solution, is
\be A_\mu^a = \varphi k_\mu \tilde c^a. \ee
From here on we switch to spheroidal coordinates, defined by
\be \ba x & = \sqrt{r^2 + a^2} \sin\theta \cos\phi \\
y & = \sqrt{r^2 + a^2} \sin\theta \sin\phi \\
z & = r \cos\theta, \ea \ee
in which the flat metric reads
\be ds^2 = -dt^2 + \frac{r^2 + a^2 \cos^2\theta}{r^2 + a^2} dr^2 + (r^2 + a^2\cos^2\theta) d\theta^2 + (a^2 + r^2)\sin^2\theta d\phi^2. \ee
In these coordinates the $\sqrt{\rm Kerr}$ gauge field is
\be \begin{array}{lr} A_t^a = \frac{r \, \tilde c^a}{r^2 + a^2 \cos^2\theta} & A_r^a = \frac{r \, \tilde c^a}{r^2 + a^2} \\ \\ A_\phi^a = -\frac{r \, \tilde c^a}{r^2 + a^2 \cos^2\theta} a\sin^2\theta \quad & A_\theta^a = 0. \end{array} \ee

\section{Review of Geodesic Double Copy}
\label{sec:GonzoReview}

In this section we review the double copy relation of test particles moving in double-copy-related backgrounds, as introduced in \cite{Gonzo:2021drq}. There are several standard Lagrangians that describe the motion of a relativistic point particle in a gravitational background. The one we find most convenient is
\be L_{\rm grav} = \half g_{\mu\nu} \dot x^\mu \dot x^\nu, \ee
where $\dot x^\mu = \frac{dx^\mu}{d\sig}$ with $\sig$ as our time parameter. The equation of motion is the geodesic equation,
\be \ba 0 & = \frac{D}{d\sig} \dot x^\mu \\
& = \ddot x^\mu + \Gamma^\mu_{\nu\rho} \dot x^\nu \dot x^\rho, \ea \ee
which in particular implies that $\sig$ is affine. The momenta are
\be \label{eq:p_grav} \ba p_\mu & = \frac{\p L_{\rm grav}}{\p \dot x^\mu} \\
& = g_{\mu\nu} \dot x^\nu, \ea \ee
with corresponding Hamiltonian
\be \ba H_{\rm grav} & = p_\mu \dot x^\mu - L_{\rm grav} \\
& = \half g^{\mu\nu} p_\mu p_\nu. \ea \ee
The Hamiltonian is conserved, and its value determines the mass via $p^2 = -m^2$. Note then that $\sig$ is related to proper time as $\tau = m\sig$. A Killing vector $\xi^\mu$ of the spacetime implies a symmetry of the Lagrangian, and by Noether's theorem there is a corresponding conserved charge
\be \ba Q_\xi & = \frac{\p L_{\rm grav}}{\p \dot x^\mu} \xi^\mu \\
& = p_\mu \xi^\mu. \ea \ee
The analogous Lagrangian for a relativistic particle with color charge moving in a gauge background is \cite{Balachandran:1976ya, Balachandran:1977ub}
\be L_{\rm gauge} = \half \bar g_{\mu\nu} \dot x^\mu \dot x^\nu - i\dot\psi^\dag \psi + c^a A_\mu^a \dot x^\mu \ee
where $\bar g_{\mu\nu}$ is the flat metric, we have set the gauge coupling to unity, $\psi$ is the particle's color vector which is valued in the fundamental representation, and we have defined
\be c^a \equiv \psi^\dag T^a \psi \ee
where $\{T^a\}$ is an orthonormal basis for the adjoint representation. The equations of motion for $\psi, \psi^\dag$ can be combined to give
\be \dot c^a = f^{abc} c^b A_\mu^c \dot x^\mu. \ee
The equation of motion for $x^\mu$, sometimes called Wong's equation \cite{Wong:1970fu}, is
\be \label{eq:lorentz} \frac{\bar D}{d\sig} \dot x^\mu = c^a F^{a,\mu}{}_\nu \dot x^\nu \ee
where $\frac{\bar D}{d\sig}$ is the covariant time derivative with respect to $\bar g_{\mu\nu}$ and $F_{\mu\nu}^a$ is the non-abelian field strength.
This can be interpreted as a relativistic and non-abelian version of the Lorentz force law. The momenta are
\be \label{eq:p_gauge} \ba p_\mu & = \frac{\p L_{\rm gauge}}{\p \dot x^\mu} \\
& = \bar g_{\mu\nu} \dot x^\nu + c^a A_\mu^a \ea \ee
and
\be p_{\psi^\dag} = \frac{\p L_{\rm gauge}}{\p \dot\psi^\dag} = -i\psi, \qquad p_\psi = \frac{\p L_{\rm gauge}}{\p \dot \psi} = 0. \ee
The Hamiltonian is then
\be \ba H_{\rm gauge} & = p_\mu \dot x^\mu + \dot\psi^\dag p_{\psi^\dag} + p_\psi \dot\psi - L_{\rm gauge} \\
& = \half \bar g_{\mu\nu} \dot x^\mu \dot x^\nu \\
& = \half \bar g^{\mu\nu} (p_\mu - c^a A_\mu^a)(p_\nu - c^b A_\nu^b). \ea \ee
Once again the conserved Hamiltonian determines the mass as $H = -\half m^2$. If $\xi^\mu$ is a Killing vector of $\bar g_{\mu\nu}$ and its Lie derivative also annihilates the gauge field, $\mathcal{L}_\xi A_\mu^a = 0$, then it implies a symmetry of the Lagrangian with Noether charge
\be \ba Q_\xi & = \frac{\p L_{\rm gauge}}{\p \dot x^\mu} \xi^\mu \\
& = p_\mu \xi^\mu. \ea \ee
When we specialize to a stationary Kerr-Schild solution and its corresponding gauge solution, as in \eqref{eq:dblcopy}, the conserved combination
\be \label{eq:bigC} C \equiv c^a \tilde c^a \ee
plays the role of an effective abelian charge and the point particle momenta \eqref{eq:p_grav} and \eqref{eq:p_gauge} can be written as
\be \ba \text{Gravity:} \qquad & p_\mu = \bar g_{\mu\nu} \dot x^\nu + \varphi k_\mu k_\nu \dot x^\nu \\
\text{Gauge:} \qquad & p_\mu = \bar g_{\mu\nu} \dot x^\nu + \varphi k_\mu \tilde c^a c^a. \ea \ee
It was observed in \cite{Gonzo:2021drq} that if the double copy relation \eqref{eq:pith} is extended as
\be \label{eq:DCext} k_\nu \,\, \longleftrightarrow \,\, \tilde c^a, \qquad \dot x^\nu \,\, \longleftrightarrow \,\, c^a, \ee
then the momenta $p_\mu$ of the two theories, and in particular the Killing charges $Q_\xi = p_\mu \xi^\mu$, are mapped to each other. Specializing further to the Kerr and $\sqrt{\rm Kerr}$ backgrounds described in section \ref{sec:KSreview}, the authors of \cite{Gonzo:2021drq} showed that for equatorial orbits the two conserved charges of energy and angular momentum are enough to determine the trajectory completely, giving a double copy relation between families of trajectories in the Kerr and $\sqrt{\rm Kerr}$ backgrounds. This suggests a deep relationship between the two point particle theories, although seemingly not a complete duality since the Hamiltonian expressions in terms of the momenta are not identical. However we will see that the correspondence goes beyond mere Killing vectors and equatorial orbits.

\section{Review of Carter Constant and Integrability in Kerr}
\label{sec:CarterReview}

The energy and angular momentum are famously not the only conserved quantities for a point particle moving in a Kerr background. There is also the Carter constant, which we review in this section along with the corresponding integrability properties.

The Carter constant \cite{Carter:1968rr} is
\be k_{\rm grav} = p_\theta^2 + a^2 m^2 \cos^2\theta + \left( \frac{p_\phi}{\sin\theta} + a\, p_t \sin\theta \right)^2. \ee
Recalling that $m^2 = -p^2$, we see it is a homogeneous quadratic polynomial in momenta and therefore can be written as
\be k_{\rm grav} = K_{\mu\nu} p^\mu p^\nu \ee
for some symmetric tensor $K_{\mu\nu}$. Let us investigate the properties of $K_{\mu\nu}$, given that $k_{\rm grav}$ is conserved. We have
\be \ba \dot k_{\rm grav} & = p^\rho \nabla_\rho \left( K_{\mu\nu} p^\mu p^\nu \right) \\
& = p^\rho p^\mu p^\nu \nabla_\rho K_{\mu\nu} \\
& = p^\rho p^\mu p^\nu \nabla_{(\rho} K_{\mu\nu)}. \ea \ee
This must vanish for all $p^\mu$, meaning that
\be \nabla_{(\rho} K_{\mu\nu)} = 0. \ee
This is the defining equation for a Killing tensor. Usually one shows that a Killing tensor implies a conserved quantity, but here we found the converse to be more instructive. More generally one can ask about the conservation of any polynomial in momenta \cite{vanHolten:2006xq}, i.e.
\be 0 = \frac{D}{d\sig} \sum_i K^{(i)}_{\mu_1\dots\mu_i} p^{\mu_1} \dots p^{\mu_i}. \ee
One finds that each of the symmetric tensors $K^{(i)}_{\mu_1\dots\mu_i}$ must be a Killing tensor. In a general theory the tensors of different rank can mix, and one gets more complicated conservation conditions. We will see this in the $\sqrt{\rm Kerr}$ theory.

The Liouville-Arnold theorem states that a Hamiltonian system with $n$ degrees of freedom and $n$ independent Poisson-commuting conserved quantities is integrable. Our point particle has four degrees of freedom $x^\mu$, and four independent conserved quantities which we can take as $H, p_t, p_\phi, k_{\rm grav}$. They also Poisson-commute, as one can quickly check using $\{x^\mu, p_\nu\} = \delta^\mu{}_\nu$. The resulting integrability is what underlies the well-known reduction to quadratures of the geodesic equation in Kerr, as reviewed for example in \cite{Kapec:2019hro}.

\section{Hidden Symmetry in $\sqrt{\rm Kerr}$}
\label{sec:HidSym}

The extended double copy relation \eqref{eq:DCext} maps $p_\mu$ in the Kerr background theory to $p_\mu$ in the $\sqrt{\rm Kerr}$ background theory. This suggests that the expression for the Carter constant, reinterpreted with $p_\mu$ as the $\sqrt{\rm Kerr}$ momentum, might be conserved in the $\sqrt{\rm Kerr}$ background theory. Indeed, conservation of
\be k_{\rm gauge} \equiv p_\theta^2 + a^2 m^2 \cos^2\theta + \left( \frac{p_\phi}{\sin\theta} + a\, p_t \sin\theta \right)^2 \ee
can be shown using eqs. \eqref{eq:bigC}, \eqref{eq:lorentz}, and \eqref{eq:p_gauge}. As in the Kerr case, this constitutes a fourth independent, Poisson-commuting constant of motion and leads to integrability for the $x^\mu$ degrees of freedom.\footnote{The gauge system technically has more than four degrees of freedom due to the color variables, but their effective abelian nature means that they can be analyzed separately from the $x^\mu$ equations. Said differently, we could just treat $c^a A_\mu^a$ as an abelian background and solve for the motion of an electric charge.} Consequently the equation of motion can be solved by quadratures, as we now show. Our derivation closely parallels that of \cite{Kapec:2019hro} for Kerr.

The first step in reducing the equation of motion to quadratures is to write the momenta in terms of the four constants of motion $p_t$, $p_\phi$, $m$, and $k_{\rm gauge}$ (along with two independent signs $\pm_r$ and $\pm_\theta$). For $p_t$ and $p_\phi$ this is tautological. Overall we find
\be p_\mu dx^\mu = p_t dt + \frac{C r \pm_r \sqrt{\mathcal{R}(r)}}{r^2 + a^2} dr \pm_\theta \sqrt{\Theta(\theta)} d\theta + p_\phi d\phi \ee
where we have defined
\be \mathcal{R}(r) \equiv \left[ (r^2 + a^2) p_t + a p_\phi \right]^2 - (r^2 + a^2) (k_{\rm gauge} + m^2 r^2) - C r \left[ 2p_t r^2 - C r + 2a(a p_t + p_\phi) \right] \ee
and
\be \Theta(\theta) \equiv k_{\rm gauge} - a^2 m^2 \cos^2\theta - \left( \frac{p_\phi}{\sin\theta} + a\, p_t \sin\theta \right)^2. \ee
Just as in the Kerr case $p_\theta$ depends only on $\theta$ and there are substantial cancellations resulting in $p_r$ depending only on $r$. The next step is to rewrite the momenta in terms of velocities, which gives
\be \ba \Sigma \, \frac{dt}{d\sig} & = -p_t r^2 + C r - a^2 p_t \cos^2\theta \\
\Sigma \, \frac{dr}{d\sig} & = \pm_r \sqrt{\mathcal{R}(r)} \\
\Sigma \, \frac{d\theta}{d\sig} & = \pm_\theta \sqrt{\Theta(\theta)} \\
\Sigma \, \frac{d\phi}{d\sig} & = \frac{C a r - a^2 p_\phi}{r^2 + a^2} + \frac{p_\phi}{\sin^2\theta} \ea \ee
where we have defined
\be \Sigma \equiv r^2 + a^2 \cos^2\theta. \ee
The signs $\pm_r$ and $\pm_\theta$ are taken to match those of $\frac{dr}{d\sig}$ and $\frac{d\theta}{d\sig}$. Next we note
\be \frac{1}{\pm_r \sqrt{\mathcal{R}(r)}} \frac{dr}{d\sig} = \frac{1}{\Sigma} = \frac{1}{\pm_\theta \sqrt{\Theta(\theta)}} \frac{d\theta}{d\sig}. \ee
Integrating this gives a relation between $r$ and $\theta$. More specifically, if our initial position is $(t_i, r_i, \theta_i, \phi_i)$ then the final coordinates $r_f, \theta_f$ will be related by
\be \fint_{r_i}^{r_f} \frac{dr}{\pm_r \sqrt{\mathcal{R}(r)}} = \fint_{\theta_i}^{\theta_f} \frac{d\theta}{\pm_\theta \sqrt{\Theta(\theta)}}. \ee
The notation $\fint$ indicates that the integrals are taken along the trajectory, possibly over multiple oscillations in $r, \theta$. Note that the integrands are always positive, so the integrals grow monotonically as they proceed along the particle's trajectory. For $t_f$ and $\phi_f$ we can use a similar trick. We write
\be \ba t_f - t_i & = \fint_{t_i}^{t_f} dt \\
& = \int_{\sig_i}^{\sig_f} \frac{dt}{d\sig} d\sig \\
& = \int_{\sig_i}^{\sig_f} \left( -p_t r^2 + C r - a^2 p_t \cos^2\theta \right) \frac{d\sig}{\Sigma} \\
& = \fint_{r_i}^{r_f} \left( -p_t r^2 + C r \right) \frac{dr}{\pm_r \sqrt{\mathcal{R}(r)}} + \fint_{\theta_i}^{\theta_f} \left( -a^2 p_t \cos^2\theta \right) \frac{d\theta}{\pm_\theta \sqrt{\Theta(\theta)}}. \ea \ee
For $\phi$ we have
\be \ba \phi_f - \phi_i & = \fint_{\phi_i}^{\phi_f} d\phi \\
& = \int_{\sig_i}^{\sig_f} \frac{d\phi}{d\sig} d\sig \\
& = \int_{\sig_i}^{\sig_f} \left( \frac{C a r - a^2 p_\phi}{r^2 + a^2} + \frac{p_\phi}{\sin^2\theta} \right) \frac{d\sig}{\Sigma} \\
& = \fint_{r_i}^{r_f} \left( \frac{C a r - a^2 p_\phi}{r^2 + a^2} \right) \frac{dr}{\pm_r \sqrt{\mathcal{R}(r)}} + \fint_{\theta_i}^{\theta_f} \left( \frac{p_\phi}{\sin^2\theta} \right) \frac{d\theta}{\pm_\theta \sqrt{\Theta(\theta)}}. \ea \ee
In summary, given the initial position $(t_i, r_i, \theta_i, \phi_i)$, the final position at time $t_f$ is completely determined by the definite integrals above. The conceptual steps in this derivation were identical to those for Kerr.

As in the gravitational case, we can view $k_{\rm gauge}$ as a quadratic polynomial in momenta. However now $m^2 = -\dot x^2 \ne -p^2$, so it will not be homogeneous. Let us explore the consequences of conservation for a general quadratic polynomial in momentum,
\be Q = K_{\mu\nu}^{(2)} p^\mu p^\nu + K_\mu^{(1)} p^\mu + K^{(0)}. \ee
The equation of motion in terms of momenta is
\be \ba \frac{\bar D}{d\sig} p^\mu & = \frac{\bar D}{d\sig} \left( \dot x^\mu + c^a A^{a,\mu} \right) \\
& = c^a \dot x^\nu \bar\nabla^\mu A_\nu^a. \ea \ee
Using this, the time derivatives of the individual terms in $Q$ are
\be \ba \frac{\bar D}{d\sig} K^{(0)} & = \dot x^\mu \bar\nabla_\mu K^{(0)} \\
\frac{\bar D}{d\sig} \left( K_\mu^{(1)} p^\mu \right) & = \dot x^\mu \dot x^\nu \bar\nabla_\nu K_\mu^{(1)} + c^a \dot x^\mu \mathcal{L}_{K^{(1)}} A_\mu^a \\
\frac{\bar D}{d\sig} \left( K_{\mu\nu}^{(2)} p^\mu p^\nu \right) & = \dot x^\rho \dot x^\mu \dot x^\nu \bar\nabla_\rho K_{\mu\nu}^{(2)} + 2c^a \dot x^\mu \dot x^\rho (A^{a,\nu} \bar\nabla_\rho K_{\mu\nu}^{(2)} + K_{\mu\nu}^{(2)} \bar\nabla^\nu A_\rho^a) \\
& \hspace{28.5mm} + c^a c^b A^{a,\mu} \dot x^\rho (A^{b,\nu} \bar\nabla_\rho K_{\mu\nu}^{(2)} + 2K_{\mu\nu}^{(2)} \bar\nabla^\nu A_\rho^b). \ea \ee
Here $\mathcal{L}_{K^{(1)}} A_\mu^a = K^{(1),\nu} \p_\nu A_\mu^a + \p_\mu K^{(1),\nu} A_\nu^a$ is the Lie derivative with respect to the vector field $K^{(1),\nu}$. If $Q$ is to be conserved then the sum of these must vanish. The vanishing of the $\mathcal{O}(\dot x^3)$ terms implies that $K_{\mu\nu}^{(2)}$ is a Killing tensor,
\be \label{eq:K2cond} 0 = \bar\nabla_{(\rho} K_{\mu\nu)}^{(2)}. \ee
This was also necessary for conservation in the gravitational case, but here it is not sufficient. We also have the vanishing of the $\mathcal{O}(\dot x^2)$ terms, which requires
\be \label{eq:K1cond} 0 = \bar\nabla_{(\mu} K_{\nu)}^{(1)} + 2c^a (A^{a,\rho} \bar\nabla_{(\mu} K_{\nu)\rho}^{(2)} + K_{\rho(\mu}^{(2)} \bar\nabla^\rho A_{\nu)}^a). \ee
If the gauge field vanished then we would have the usual Killing vector condition on $K_\mu^{(1)}$, but here we have two additional terms involving $K_{\mu\nu}^{(2)}$. Finally the vanishing of the $\mathcal{O}(\dot x)$ terms requires
\be \label{eq:K0cond} 0 = \bar\nabla_\mu K^{(0)} + c^a \mathcal{L}_{K^{(1)}} A_\mu^a + c^a c^b A^{a,\rho} (A^{b,\nu} \bar\nabla_\mu K_{\rho\nu}^{(2)} + 2K_{\rho\nu}^{(2)} \bar\nabla^\nu A_\mu^b) \ee
which involves all three of $K_{\mu\nu}^{(2)}$, $K_\mu^{(1)}$, and $K^{(0)}$. These equations describe the gauge analogue of a rank two Killing tensor. For $k_{\rm gauge}$ we have
\be k_{\rm gauge} = K_{\mu\nu}^{(2)} p^\mu p^\nu + K_\mu^{(1)} p^\mu + K^{(0)} \ee
with
\be \ba K^{(2)}_{\mu\nu} & = (\p_\theta)_\mu (\p_\theta)_\nu - a^2 \bar g_{\mu\nu} \cos^2\theta + \left( \frac{\p_\phi}{\sin\theta} + a\sin\theta \, \p_t \right)_\mu \left( \frac{\p_\phi}{\sin\theta} + a\sin\theta \, \p_t \right)_\nu \\
K^{(1)}_\mu & = 2a^2 c^a A_\mu^a \cos^2\theta \\
K^{(0)} & = -a^2 (c^a A_\mu^a) (c^b A^{b,\mu}) \cos^2\theta = 0 \ea \ee
where $(\p_\mu)^\nu = \delta^\nu_\mu$ are the coordinate vector field components. One can check that eqs. \eqref{eq:K2cond}-\eqref{eq:K0cond} are satisfied. This perspective is especially interesting because it emphasizes special properties of the $\sqrt{\rm Kerr}$ background itself, as opposed to merely the particle moving on the background. It suggests that the double copy respects hidden symmetries.

\section{Discussion}
\label{sec:Discussion}

After reviewing the Kerr-Schild double copy and its extension to include point particles moving on the respective backgrounds, we established the existence of a single copy analogue of the Carter constant for the $\sqrt{\rm Kerr}$ solution. It led to integrability of the equations of motion of a point particle, and we showed how to reduce them to quadratures. We then showed how to reinterpret this constant of motion as a geometric statement about the background itself, showing that, at least for the Kerr spacetime, the double copy preserves hidden symmetries. One may wonder to what extent the more powerful Weyl double copy \cite{Luna:2018dpt} makes this manifest. After all, the Weyl double copy is built from the rank-two Killing spinor possessed by any Petrov type D spacetime \cite{Walker:1970un}. However, this Killing spinor is not necessarily associated with any Killing tensor, but rather a conformal Killing tensor. Closely related to this, it leads to a constant of motion only for massless particles, namely the Penrose-Walker constant. The Kerr spacetime's Killing tensor is related to its Killing spinor, but somewhat nontrivially; it is not simply a special case of the associated conformal Killing tensor. Therefore, while it is not shocking, neither is it obvious to us why the double copy should respect the hidden symmetry of Kerr. Our results do seem related to those of \cite{Chawla:2022ogv} though, where it was found that the single copy gauge field corresponding to a Kerr-NUT-(A)dS spacetime \cite{Carter:1968rr} has a field strength that is ``aligned'' \cite{Krtous:2007xg} with (but not proportional to) the spacetime's principal tensor \cite{Kubiznak:2006kt}. The principal tensor controls the Killing tensors of Kerr-NUT-(A)dS, as reviewed in \cite{Frolov:2017kze}, so their results provide further evidence that the double copy somehow respects hidden symmetries. In any case, it seems likely that our results generalize beyond Kerr, perhaps even to the entire Kerr-NUT-(A)dS family. It will be very interesting to see how far they extend.

\acknowledgments

We are grateful to Tucker Manton and Marcus Spradlin for useful discussions, and to Cynthia Keeler for comments on the draft. This work was supported in part by the US Department of Energy under contract DE-SC0010010 Task F and by Simons Investigator Award \#376208. Y. Chen was also supported by a Karen T. Romer Undergraduate Teaching and Research Award.

\bibliography{main}
\bibliographystyle{JHEP}

\end{document}